\documentclass[12pt]{article}
\usepackage{draft,tikz-cd,shuffle}
\def\ma{\mathcal}

\begin{document}

\begin{titlepage}

\begin{center}

\hfill \\
\hfill \\
\vskip 1cm

\title{Berends-Giele currents for extended gravity}

\bigskip

\author{Yi-Xiao Tao$^{a}$}

\bigskip

\address{${}^a$Department of Mathematical Sciences, Tsinghua University, Beijing 100084, China}

\email{taoyx21@mails.tsinghua.edu.cn}

\end{center}

\vfill

\begin{abstract}
In this short paper, we write down the Berends-Giele (BG) currents for extended gravity explicitly and discuss the unifying relations of these BG currents. This new tool, different from the double field theory current formally, may deepen our understanding of the current Kawai-Lewellen-Tye (KLT) relation.

\end{abstract}

\vfill

\end{titlepage}

\tableofcontents

\newpage

\section{Introduction}
Gravity is still the most mysterious thing in our nature. The exploration of a quantum theory of gravity never stops. The first trial of quantum gravity is the perturbative gravity theory \cite{DeWitt:1967ub,DeWitt:1967uc,DeWitt:1967yk}, which regards the gravitons as the fluctuation of the flat metric. In this way, we can obtain the Lagrangian of the pure graviton theory by evaluating the Einstein-Hilbert action. The derivation of the Feynman rules and the amplitudes are straightforward. After the amazing discovery of duality between the Anti-de Sitter gravity and the Conformal field theory (AdS/CFT) \cite{Maldacena:1997re,Witten:1998qj}, we began to learn how to deal with the non-perturbative gravity theory from the field theory perspective. 

For a more general gravity theory, say extended gravity theory, we also include the antisymmetric B-fields and dilatons. The theory, as a generalization of the traditional gravity, comes from the effective action of string theory. This theory is also the result of the double copy prescription of the Yang-Mills (YM) theory \cite{Bern:2019prr}. Our knowledge of the double copy between the YM amplitudes and the extended gravity amplitudes is very rich. On the contrary, the ``double copy" property for the off-shell objects of both sides is still not clear. In addition, many relations for amplitudes do not hold exactly for off-shell objects. Hence to see how these amplitude relations are realized on off-shell objects will be interesting. In this paper, we will focus on the simplest off-shell object, Berends-Giele (BG) currents, which can be defined as amplitudes with one leg off-shell \cite{Berends:1987me}. BG currents also have a nice recursion property, and the recursion relation can be obtained from the classical equation of motion. This property makes BG currents easy to generalize to different theories and even different background metrics as long as we can write down the classical equation of motion of those theories. An equivalent way to obtain the gravity currents is using the double field theory \cite{Cho:2021nim}. In \cite{Cho:2021nim}, they discuss the current KLT relation, as a generalization of the KLT relation for amplitudes \cite{Kawai:1985xq,Bern:2010yg}, for pure graviton currents and write down the explicit formalism for some low-point examples. Our work can be regarded as a complement to this nice work\footnote{For the double field theory method, we consider the perturbative expansion of the generalized metric \cite{Cho:2021nim}. If we consider the B-field and the dilaton in the double field theory, we will obtain an equivalent result of the extended gravity currents.}. In addition, some double copy relations for effective theory currents have been studied in \cite{Mizera:2018jbh}.

In this paper, we will write down the BG currents of extended gravity explicitly and explore some properties of them. Then we will consider the unifying relation for amplitudes and finally find that this relation does not hold exactly for gravity BG currents. The paper will be organized as follows. In section \ref{sec2}, we review how to obtain the BG currents of the pure graviton theory. In section \ref{sec3}, we write down the Lagrangian of the extended gravity and derive the BG currents of gravitons, B-fields, and dilatons. In section \ref{sec4}, we derive ``off-shell" unifying relations for 2-pt, 3-pt, and 4-pt currents\footnote{The word ``$n$-pt" here has the same meaning as ``rank-n" in \cite{Cho:2021nim}.} to fix the parameter of the theory and to find some hints of the off-shell double copy relation.

\section{Review: BG currents for pure gravitons}\label{sec2}
In this section, we will review how to obtain BG currents for pure gravitons. We will follow the process in \cite{Gomez:2021shh}. More discussion about pure graviton currents can be seen in \cite{Cho:2022faq}. Let us consider the following Einstein-Hilbert action:
\ie
S_{\text{EH}}=\frac{1}{2\kappa^2}\int d^D x \sqrt{-g}R
\fe
where $g$ is the determinant of the metric $g_{\mu\nu}$ and $R$ is the corresponding scalar curvature. In the following discussion, we will choose $\kappa=1$. The equation of motion (EOM) can be found from the condition $\delta S/\delta g^{\mu\nu}=0$:
\ie
R_{\mu\nu}-\frac{1}{2}g_{\mu\nu}R=0
\fe
where $R_{\mu\nu}$ is the Ricci tensor and $g^{\mu\nu}R_{\mu\nu}=R$. For $D>2$, we will find $R=0$ and the EOM can be simplified as 
\ie
R_{\mu\nu}=0.
\fe
Now we consider the following perturbiner ansatz:
\ie\label{ansatz1}
&g_{\mu\nu}=\eta_{\mu\nu}+\sum_{I}H_{I\mu\nu}e^{ik_{I}\cdot x}\\
&g^{\mu\nu}=\eta^{\mu\nu}-\sum_{I}F_I^{\mu\nu}e^{ik_{I}\cdot x}
\fe
where we always choose $I$ to be the order of the natural numbers. From $g_{\mu\nu}g^{\nu\gamma}=\delta_{\mu}^{\gamma}$, we have
\ie\label{inv}
F_{I}^{\mu\nu}=\eta^{\mu\rho}\eta^{\nu\sigma}H_{I\rho\sigma}-\eta^{\nu\sigma}\sum_{I=X\cup Y}F_{X}^{\mu\rho}H_{Y\rho\sigma}
\fe
where $I=X\cup Y$ means that $I\in X\shuffle Y$\footnote{Some examples can be seen in \cite{Mizera:2018jbh}.}. For example, if $I=123$, then $\sum_{I=X\cup Y}$ means that we need to sum over these terms: $(X,Y)=(1,23),(12,3),(13,2),(23,1),(3,12),(2,13)$. The Riemann curvature tensor can be obtained by the following equation:
\ie
R_{\mu\rho\nu}^{~~~~\sigma}=\partial_{\rho}\Gamma^{\sigma}_{\mu\nu}-\partial_{\nu}\Gamma^{\sigma}_{\mu\rho}+\Gamma^{\sigma}_{\rho\lambda}\Gamma^{\lambda}_{\mu\nu}-\Gamma^{\sigma}_{\nu\lambda}\Gamma^{\lambda}_{\mu\rho}
\fe
and $R_{\mu\nu}=R_{\mu\rho\nu}^{~~~\rho}$. The Christoffel symbol $\Gamma^{\sigma}_{\mu\nu}=g^{\sigma\rho}\Gamma_{\mu\nu\rho}$ is given by
\ie
\Gamma_{\mu\nu\rho}=\partial_{\mu}g_{\nu\rho}+\partial_{\nu}g_{\mu\rho}-\partial_{\rho}g_{\mu\nu}
\fe
Substituting \eqref{ansatz1} into the EOM and choosing the gauge $\eta^{\mu\nu}\Gamma_{\mu\nu\rho}=0$ which corresponds to
\ie
\eta^{\mu\nu}(k_{I\mu}H_{I\nu\rho}-\frac{1}{2}k_{I\rho}H_{I\mu\nu})=0,
\fe
we will find that
\ie
-s_{I}H_{I\mu\nu}=&-2i\sum_{I=X\cup Y}F^{\rho\sigma}_{X}(k_{I\rho}\Gamma_{Y\mu\nu\sigma}-k_{I\nu}\Gamma_{Y\mu\rho\sigma})-2\eta^{\alpha\beta}\eta^{\rho\sigma}\sum_{I=X\cup Y}(\Gamma_{X\nu\alpha\sigma}\Gamma_{Y\mu\rho\beta}-\Gamma_{X\rho\alpha\sigma}\Gamma_{Y\mu\nu\beta})\\
&+2\eta^{\alpha\beta}\sum_{I=X\cup Y\cup Z}F^{\rho\sigma}_{X}(\Gamma_{Y\nu\rho\beta}\Gamma_{Z\mu\alpha\sigma}-\Gamma_{Y\alpha\rho\beta}\Gamma_{Z\mu\nu\sigma}+\Gamma_{Y\nu\alpha\sigma}\Gamma_{Z\mu\rho\beta}-\Gamma_{Y\rho\alpha\sigma}\Gamma_{Z\mu\nu\beta})\\
&-2\sum_{I=X\cup Y\cup Z\cup W}F_{X}^{\rho\sigma}F_{Y}^{\alpha\beta}(\Gamma_{Z\nu\alpha\sigma}\Gamma_{W\mu\rho\beta}-\Gamma_{Z\rho\alpha\sigma}\Gamma_{W\mu\nu\beta})
\fe
with
\ie
\Gamma_{I\mu\nu\rho}=\frac{i}{2}(k_{I\mu}H_{I\nu\rho}+k_{I\nu}H_{I\mu\rho}-k_{I\rho}H_{I\mu\nu}).
\fe
Here $s_{I}=k_{I}^2=(\sum_{i\in I}k_{i})^2$. Finally, we need to give the single-particle states. The single-particle states of gravitons are symmetric tensors:
\ie
H_{i\mu\nu}=h_{i\mu\nu}
\fe
with $\eta^{\mu\nu}h_{i\mu\nu}=\eta^{\mu\nu}k_{\mu}h_{i\nu\rho}=0$ which are the traceless condition in the flat spacetime and the transverse condition respectively. The amplitude for pure gravitons is given by
\ie
M(In)=\lim_{k_{n}^2\to0}H_{Iab}h_{ncd}\eta^{ac}\eta^{bd}.
\fe

\section{BG currents of extended gravity}\label{sec3}
In this section, we will consider the extended gravity theory. Firstly we write down the action of extended gravity, which is the effective theory of string theory \cite{Bern:2019prr,Gross:1986mw}:
\ie
S_{\text{EG}}=\int d^D x \sqrt{-g}\bigg[-\frac{1}{2}R+\frac{1}{2(D-2)}g^{\mu\nu}\partial_{\mu}\phi\partial_{\nu}\phi+\frac{1}{24}e^{-4\phi/(D-2)}g^{a\lambda}g^{b\mu}g^{c\nu}H_{abc}H_{\lambda\mu\nu}\bigg]
\fe
where
\ie
H_{\lambda\mu\nu}=\partial_{\lambda}B_{\mu\nu}+\partial_{\mu} B_{\nu\lambda}+\partial_{\nu}B_{\lambda\mu}.
\fe
Here we still use the convention $\kappa=1$. Note that the scale of the B-field here is different from \cite{Bern:2019prr}. The reason for this convention will be given in section \ref{sec4} where we find that only when the coefficient is $1/24$, the unifying relation \cite{Cheung:2017ems}, as a corollary of the KLT relation, is correct.

As usual, we can obtain the EOM from the action:
\ie
R_{\mu\nu}-\frac{1}{2}g_{\mu\nu}R=&-g_{\mu\nu}(\frac{1}{2(D-2)}g^{ab}\partial_{a}\phi\partial_{b}\phi+\frac{1}{24}e^{-4\phi/(D-2)}g^{a\lambda}g^{bk}g^{cl}H_{abc}H_{\lambda kl})+\\
&\frac{1}{D-2}\partial_{\mu}\phi\partial_{\nu}\phi+\frac{1}{4} e^{-4\phi/(D-2)}g^{ac}g^{bd}H_{\mu ab}H_{\nu cd}\\
g^{m\nu}\partial_{m}\partial_{\nu}\phi=&\frac{1}{2}g_{ab}\partial_{m}g^{ab}g^{m\nu}\partial_{\nu}\phi-\partial_{m}g^{m\nu}\partial_{\nu}\phi-\frac{1}{3}e^{-4\phi/(D-2)}g^{a\lambda}g^{b\mu}g^{c\nu}H_{abc}H_{\lambda\mu\nu}\\
e^{-4\phi/(D-2)}g^{a\lambda}g^{bm}g^{cn}\partial_{\lambda}H_{abc}\eta_{m\mu}\eta_{n\nu}=&e^{-4\phi/(D-2)}\bigg[\frac{1}{2}g_{kl}\partial_{\lambda}g^{kl}g^{a\lambda}g^{bm}g^{cn}H_{abc}\eta_{m\mu}\eta_{n\nu}\\
&+\frac{4}{D-2}\partial_{\lambda}\phi g^{a\lambda}g^{bm}g^{cn}H_{abc}\eta_{m\mu}\eta_{n\nu}-\partial_{\lambda}g^{a\lambda}g^{bm}g^{cn}H_{abc}\eta_{m\mu}\eta_{n\nu}\\
&-g^{a\lambda}\partial_{\lambda}g^{bm}g^{cn}H_{abc}\eta_{m\mu}\eta_{n\nu}-g^{a\lambda}g^{bm}\partial_{\lambda}g^{cn}H_{abc}\eta_{m\mu}\eta_{n\nu}\bigg]
\fe
It is worth mentioning that we cannot cancel the factor $e^{-4\phi/(D-2)}$ in the EOM of the B-field. Although this operation will not change the solution of the EOM, it will change BG currents which correspond to the coefficients of certain orders of the solution. To match the Feynman rules, we must keep this factor and this factor will give contact terms with any number of dilatons. For example, if we only focus on the 4-vertex $\phi\phi BB$, then we can truncate the power series of the factor $e^{-4\phi/(D-2)}$ to the term $\frac{(-4)^2}{2!(D-2)^2}\phi^2$ and then calculate the BG currents. If we cancel the factor $e^{-4\phi/(D-2)}$ on both sides, we will not obtain this 4-vertex.

To obtain BG currents, we need to expand the factor $e^{-4\kappa\phi/(D-2)}$ as a power series of $\phi$. Also note that the scalar curvature can be written as
\ie
\frac{2-D}{2}R=-\frac{1}{2}g^{ab}\partial_{a}\phi\partial_{b}\phi+\frac{6-D}{24}e^{-4\phi/(D-2)}g^{a\lambda}g^{bk}g^{cl}H_{abc}H_{\lambda kl}
\fe
After some calculations, we obtain the currents for extended gravity. The graviton currents can be found in appendix \ref{app1}.

The B-field current and the dilaton current can also be derived similarly. It is worth mentioning that, as a 2-form field, the gauge for B-fields is chosen to be
\ie
\eta^{\mu\nu}\partial_{\mu}B_{\nu\rho}=0.
\fe
The single-particle state for gravitons is the same as the pure graviton case, while the single-particle state for B-fields should be chosen to be antisymmetric and the single-particle state should be chosen to be transverse diagonal \cite{Gross:1986mw,Bern:2019prr}. Then the BG current for fat gravitons (i.e. all external legs should be the sum of single-particle states of the graviton, the B-field, and the dilaton) can be obtained by summing over the graviton current, the B-field current, and the dilaton current (with a projective operator). Then the amplitude of fat gravitons can be obtained from the BG current as before. 

As we have shown, the vertices involving dilatons make BG currents very complicated, which is also the greatest obstacle to the applications of these currents. In the next section, we will explore the applications of these currents after truncating some dilaton vertices.

\section{Unifying relation for extended gravity currents}\label{sec4}
The ultimate goal of constructing extended gravity currents is constructing the KLT relation for BG currents, i.e. finding the extra off-shell terms after imposing the traditional KLT relation to BG currents naively, which will be zero when we take the on-shell limit and obtain amplitudes. However, finding the explicit form of the current KLT relation is extremely difficult. In \cite{Cho:2021nim}, they calculate some low-point cases and we will not repeat this result in this paper. Instead, we discuss the unifying relation, as a corollary of the KLT relation, for extended gravity currents as a verification of this new tool. In this case, as we will see, the extra off-shell terms appear and the number of such terms explodes when the number of points of BG current increases. We will only construct the unifying relation and the off-shell terms in detail for 2-pt, 3-pt, and 4-pt currents in this paper.

\subsection{Unifying relation for extended gravity}
It is interesting to explore the off-shell version of the unifying relations. In \cite{Cheung:2017ems}, the unifying relation for extended gravity amplitudes has been raised, which is also a corollary of the KLT relation:
\ie
\mathcal{T}[\alpha]A_{\text{EG}}=A_{\text{YM}}(\alpha)
\fe
where the operator $\ma{T}[\alpha]$ is defined as
\ie
\ma{T}[\alpha]=\partial_{\epsilon_{\alpha_{1}}\epsilon_{\alpha_{n}}}\prod_{i=1}^{n-2}(\partial_{k_{\alpha_i}\epsilon_{\alpha_{i+1}}}-\partial_{k_{\alpha_n}\epsilon_{\alpha_{i+1}}}).
\fe
The polarization vector for extended gravity has a double copy formalism $\epsilon_{\mu}\tilde{\epsilon}_{\nu}$ and in this paper we choose the differential operator $\ma{T}[\alpha]$ to act on the $\tilde{\epsilon}$ part. This relation converts the extended gravity amplitudes into the color-ordered Yang-Mills amplitudes with the color order $\alpha$. However, for BG currents, which have 1 leg being off-shell, there may be some extra terms after acting $\ma{T}[\alpha]$ on the extended gravity currents which will vanish when we take the on-shell limit. For the unifying relation between the Yang-Mills theory and the Yang-Mills scalar theory, there is no such term \cite{Tao:2022nqc,Chen:2023bji}. However, things are different for extended gravity, as we will see.

From the definition of $\ma{T}[\alpha]$, we will find that the terms that survive must not contain dilatons since the dilaton propagators do not contain $\eta_{\mu\nu}$. Thus we can truncate all dilaton terms:
\ie
R_{\mu\nu}=&\frac{1}{6(2-D)}g_{\mu\nu}g^{a\lambda}g^{bk}g^{cl}H_{abc}H_{\lambda kl}+\frac{1}{4}g^{ac}g^{bd}H_{\mu ab}H_{\nu cd}\\
g^{a\lambda}g^{bm}g^{cn}\partial_{\lambda}H_{abc}\eta_{m\mu}\eta_{n\nu}&=\frac{1}{2}g_{kl}\partial_{\lambda}g^{kl}g^{a\lambda}g^{bm}g^{cn}H_{abc}\eta_{m\mu}\eta_{n\nu}-\partial_{\lambda}g^{a\lambda}g^{bm}g^{cn}H_{abc}\eta_{m\mu}\eta_{n\nu}\\
&-g^{a\lambda}\partial_{\lambda}g^{bm}g^{cn}H_{abc}\eta_{m\mu}\eta_{n\nu}-g^{a\lambda}g^{bm}\partial_{\lambda}g^{cn}H_{abc}\eta_{m\mu}\eta_{n\nu}
\fe
For single-particle states, we will use the following double-copy prescription
\ie
H_{i\mu\nu}&=\frac{1}{2}(\epsilon_{\mu}\tilde{\epsilon}_{\nu}+\epsilon_{\nu}\tilde{\epsilon}_{\mu})\\
B_{i\mu\nu}&=\frac{1}{2}(\epsilon_{\mu}\tilde{\epsilon}_{\nu}-\epsilon_{\nu}\tilde{\epsilon}_{\mu})
\fe
with $\epsilon_{i}\cdot\tilde{\epsilon}_{i}=0$, which means that we will not consider dilaton external legs. To make this paper self-contained, we write down the gluon BG current with Lorenz gauge here:
\ie
s_{I}A_{I\mu}=&\sum_{I=XY}(k_{X}-k_{Y})_{\mu}(A_{X}\cdot A_{Y})-2A_{X\mu}(k_{X}\cdot A_{Y})+2A_{Y\mu}(k_{Y}\cdot A_{X})\\
&+\sum_{I=XYZ}2A_{Y\mu}A_{X}\cdot A_{Z}-A_{X\mu}A_{Y}\cdot A_{Z}-A_{Z\mu}A_{X}\cdot A_{Y}.
\fe
The reason for choosing the Lorenz gauge here is that if we act the operator $\ma{T}[\alpha]$ on the gauge condition for graviton currents we will expect to obtain gluon currents with the Lorenz gauge. We will expect the following equation:
\ie\label{exp}
\ma{T}[In](H_{Iab}+B_{Iab})\epsilon_{c}\tilde{\epsilon}_{d}\eta^{ac}\eta^{bd}\sim A_{I}\cdot\epsilon_{n}+\text{extra off-shell terms}.
\fe

\subsection{2-pt current}
Let us start with the simplest case, 2-pt currents (i.e. $I=12$), which only involves the 3-vertex terms. We can find out such terms:
\ie
-s_{I}H_{I\mu\nu}\supset&-2i\sum_{I=X\cup Y}F^{\rho\sigma}_{X}(k_{I\rho}\Gamma_{Y\mu\nu\sigma}-k_{I\nu}\Gamma_{Y\mu\rho\sigma})-2\eta^{\alpha\beta}\eta^{\rho\sigma}\sum_{I=X\cup Y}(\Gamma_{X\nu\alpha\sigma}\Gamma_{Y\mu\rho\beta}-\Gamma_{X\rho\alpha\sigma}\Gamma_{Y\mu\nu\beta})\\
&-\frac{1}{2} \eta^{ac}\eta^{bd}\sum_{I=X\cup Y}H_{X\mu ab}H_{Y\nu cd}+\frac{1}{3(D-2)}\eta_{\mu\nu}\eta^{a\lambda}\eta^{bk}\eta^{cl}\sum_{I=X\cup Y}H_{Xabc}H_{Y\lambda kl}\\
-s_{I}B_{I\mu\nu}\supset&-i\frac{1}{2}\eta_{kl}\eta^{a\lambda}\sum_{I=X\cup Y}k_{X\lambda}F^{kl}_XH_{Ya\mu\nu}+i\sum_{I=X\cup Y}k_{X\lambda}F^{a\lambda}_XH_{Ya\mu\nu}\\
&+i\eta^{a\lambda}\sum_{I=X\cup Y}k_{X\lambda}F^{bm}_XH_{Yab\nu}\eta_{m\mu}+i\eta_{n\nu}\eta^{a\lambda}\sum_{I=X\cup Y}k_{X\lambda}F^{cn}_XH_{Ya\mu c}\\
&+i\sum_{I=X\cup Y}(F^{a\lambda}_Xk_{Y\lambda}H_{Ya\mu\nu}+\eta^{a\lambda}F^{bm}_{X}k_{Y\lambda}H_{Yab\nu}\eta_{m\mu}-\eta^{a\lambda}F^{bm}_{X}k_{Y\lambda}H_{Yab\mu}\eta_{m\nu})
\fe
From the equation
\ie
\ma{T}[12]H_{1ab}\epsilon_{2c}\tilde{\epsilon}_{2d}\eta^{ac}\eta^{bd}=\ma{T}[12]H_{1ab}\epsilon_{2c}\tilde{\epsilon}_{2d}\eta^{ac}\eta^{bd}=\frac{1}{2}\epsilon_{1}\cdot\epsilon_{2}=\frac{1}{2}A_{1}\cdot\epsilon_{2}
\fe
and the definition of the operator $\ma{T}[\alpha]$, we can find each term in the gluon currents after acting $\ma{T}[123]$ on the currents $H_{12ab}\epsilon_{3c}\tilde{\epsilon}_{3d}\eta^{ac}\eta^{bd}$ and $B_{12ab}\epsilon_{3c}\tilde{\epsilon}_{3d}\eta^{ac}\eta^{bd}$. Also note that for all $X$ and $Y$ such that $I=X\cup Y$, only the cases $I=XY$ and $I=YX$ are possible to survive from $\ma{T}[In]$.

We first investigate the pure graviton vertices in the graviton current. After acting $\ma{T}[123]$ on $H_{12ab}\epsilon_{3c}\tilde{\epsilon}_{3d}\eta^{ac}\eta^{bd}$, the first term 
\ie
\sum_{I=X\cup Y}-2iF^{\rho\sigma}_{X}k_{I\rho}\Gamma_{Y\mu\nu\sigma}
\fe
will give 
\ie
\frac{1}{4}\epsilon_{1}\cdot\epsilon_{2}k_{1}\cdot\epsilon_{3}-\frac{1}{2}\epsilon_{1}\cdot\epsilon_{3}k_{1}\cdot\epsilon_{2}
\fe
while the term proportional to $k_{I\nu}$ will not contribute since such terms will not give  $\tilde{\epsilon}_{1}\cdot\tilde{\epsilon}_{3}$. The second term 
\ie
-2\eta^{\alpha\beta}\eta^{\rho\sigma}\sum_{I=X\cup Y}(\Gamma_{X\nu\alpha\sigma}\Gamma_{Y\mu\rho\beta}-\Gamma_{X\rho\alpha\sigma}\Gamma_{Y\mu\nu\beta})
\fe
will give 
\ie
\frac{1}{4}\epsilon_{1}\cdot k_{2}\epsilon_{2}\cdot\epsilon_{3}.
\fe
Note that there is a very subtle term in it: 
\ie
\sum_{I=X\cup Y}\eta^{\alpha\beta}\eta^{\rho\sigma}k_{X\sigma}H_{X\alpha\nu}k_{Y\rho}H_{Y\mu\beta}.
\fe
This term will not contribute to the unifying relation in the 2-pt current case. However, when we move to the higher point case, we will find that $\eta^{\alpha\beta}H_{X\alpha\nu}H_{Y\mu\beta}$ has enough $k\cdot\epsilon$ and it will survive from the acting of $\ma{T}[\alpha]$.

Other terms come from the $HBB$-vertex terms. The term 
\ie
\frac{1}{2}\eta^{ac}\eta^{bd}\sum_{I=X\cup Y}(k_{X\mu}B_{Xab}+k_{Xa}B_{Xb\mu}+k_{Xb}B_{X\mu a})(k_{Y\nu}B_{Ycd}+k_{Yc}B_{Yd\nu}+k_{Yd}B_{Y\nu c})
\fe
will give 
\ie
-\frac{1}{4}\epsilon_{1}\cdot\epsilon_{2}k_{2}\cdot\epsilon_{3}+\frac{1}{4}\epsilon_{1}\cdot k_{2}\epsilon_{2}\cdot\epsilon_{3}
\fe 
and the last term, which proportional to $\eta_{\mu\nu}$ will not contribute.

Now we can sum over all terms and obtain the following equation:
\ie
-\ma{T}[123]s_{12}H_{12ab}\epsilon_{3c}\tilde{\epsilon}_{3d}\eta^{ac}\eta^{bd}=\frac{1}{4}s_{12}A_{12}\cdot\epsilon_{3}.
\fe
For the B-field currents, we can also do the same analysis. Finally we have
\ie
-\ma{T}[123]s_{12}H_{12ab}\epsilon_{3c}\tilde{\epsilon}_{3d}\eta^{ac}\eta^{bd}=-\ma{T}[123]s_{12}B_{12ab}\epsilon_{3c}\tilde{\epsilon}_{3d}\eta^{ac}\eta^{bd}=\frac{1}{4}s_{12}A_{12}\cdot\epsilon_{3}.
\fe

\subsection{3-pt current}
Let us turn to 3-pt currents. Now we need to consider the contribution from the 4-vertex terms and terms like $$\sum_{I=X\cup Y}\eta^{\alpha\beta}\eta^{\rho\sigma}k_{X\sigma}H_{X\alpha\nu}k_{Y\rho}H_{Y\mu\beta}.$$ We still only discuss the graviton currents as an example. From the results of the 2-pt currents, the 3-vertex terms of the gluon currents can be easily obtained. The 4-vertex terms of the graviton current $-s_{123}H_{123ab}\epsilon_{4c}\tilde{\epsilon}_{4d}\eta^{ac}\eta^{bd}$ will give the following terms after acting $\ma{T}[1234]$ on it:
\ie
-\frac{1}{4}\epsilon_{1}\cdot\epsilon_{3}\epsilon_{2}\cdot\epsilon_{4}+\frac{1}{8}\epsilon_{1}\cdot\epsilon_{4}\epsilon_{2}\cdot\epsilon_{3}.
\fe
Note that when we talk about the 4-vertex terms of the graviton currents, we also include the 2-deshuffle terms (terms with $\sum_{I=X\cup Y}$) with an inverse graviton subcurrent say $F_{X}^{\mu\nu}$, since from \eqref{inv}, such currents are also possible to contribute to 4-vertex terms. In fact, these terms contribute $\frac{1}{8}\epsilon_{1}\cdot\epsilon_{4}\epsilon_{2}\cdot\epsilon_{3}$.

From the gluon currents, we know we also need a term 
\ie
\frac{1}{8}\epsilon_{1}\cdot\epsilon_{2}\epsilon_{3}\cdot\epsilon_{4}.
\fe
This term comes from
\ie\label{hh}
\sum_{I=X\cup Y}\eta^{\alpha\beta}\eta^{\rho\sigma}k_{X\sigma}H_{X\alpha\nu}k_{Y\rho}H_{Y\mu\beta}=\frac{1}{2}\sum_{I=X\cup Y}(k_{I}^2-k_{X}^2-k_{Y}^2)\eta^{\alpha\beta}H_{X\alpha\nu}H_{Y\mu\beta}
\fe
and
\ie
\sum_{I=X\cup Y}\eta^{\alpha\beta}\eta^{\rho\sigma}k_{X\sigma}B_{X\alpha\nu}k_{Y\rho}B_{Y\mu\beta}=\frac{1}{2}\sum_{I=X\cup Y}(k_{I}^2-k_{X}^2-k_{Y}^2)\eta^{\alpha\beta}B_{X\alpha\nu}B_{Y\mu\beta}.
\fe
A simple calculation shows that these two terms will also give an extra off-shell term
\ie
-\frac{s_{123}}{8s_{12}}(\epsilon_{1}\cdot\epsilon_{2})(\epsilon_{3}\cdot\epsilon_{4})
\fe
which will vanish when will take $k_{4\mu}$ to be on-shell. Another extra off-shell term comes from
\ie
&\sum_{I=X\cup Y}(\eta^{a\sigma}H_{X\rho a}k_{Y}^{\rho}k_{Y\mu}H_{Y\sigma\nu}-\eta^{bd}k_{X\mu}B_{Xba}k_{Y}^aB_{Yd\nu}).
\fe
For this term, if $k\cdot\tilde{\epsilon}$ comes from factors like $k_{Y}^{\rho}H_{X\rho a}$, it will give the term which coincides with the term in the gluon currents. If $k\cdot\tilde{\epsilon}$ comes from terms like $H_{X\rho a}H_{Y\sigma\nu}\eta^{a\sigma}$, it will give an extra off-shell term
\ie
\frac{1}{8s_{12}}k_{123}\cdot\epsilon_{4}k_{12}\cdot\epsilon_{3}\epsilon_{1}\cdot\epsilon_{2}.
\fe

After a similar analysis for the B-field currents, we finally find
\ie
-\ma{T}[1234]s_{123}H_{123ab}\epsilon_{4c}\tilde{\epsilon}_{4d}\eta^{ac}\eta^{bd}&=-\frac{1}{8}A_{123}\cdot\epsilon_{4}-\frac{k_{123}^2}{8s_{12}}(\epsilon_{1}\cdot\epsilon_{2})(\epsilon_{3}\cdot\epsilon_{4})+\frac{1}{8s_{12}}k_{123}\cdot\epsilon_{4}k_{12}\cdot\epsilon_{3}\epsilon_{1}\cdot\epsilon_{2}\\
-\ma{T}[1234]s_{123}B_{123ab}\epsilon_{4c}\tilde{\epsilon}_{4d}\eta^{ac}\eta^{bd}&=-\frac{1}{8}A_{123}\cdot\epsilon_{4}-\frac{k_{123}^2}{8s_{12}}(\epsilon_{1}\cdot\epsilon_{2})(\epsilon_{3}\cdot\epsilon_{4})+\frac{1}{8s_{12}}k_{123}\cdot\epsilon_{4}k_{12}\cdot\epsilon_{3}\epsilon_{1}\cdot\epsilon_{2}
\fe
The greatest difficulty in this case is that there will also be some contribution from the 3-vertex terms. In other words, unlike the unifying relation between YM and YMS, where 4-vertex only corresponds to 4-vertex, 4-vertex in YM corresponds to both 4-vertex and two 3-vertex in the extended gravity. Note that in the 3-pt case, the extra off-shell terms appear for the first time. In the 4-pt currents case, we will find that these extra off-shell terms are necessary but hard to deal with.

\subsection{4-pt current}
The analysis is similar to the former cases. In this case, however, we need to deal with some extra terms in addition to the terms that the gluon currents consist of. Such terms will come from the off-shell terms in the 3-pt current case, terms like \eqref{hh} and 4-vertex. Note that in this case the 5-vertex will not contribute.

After acting $\ma{T}[12345]$ on $-s_{1234}H_{1234ab}\epsilon_{5c}\tilde{\epsilon}_{5d}\eta^{ac}\eta^{bd}$, the extra term from 3-vertex terms are (including the extra off-shell terms of the 3-pt currents and terms like \eqref{hh})
\ie
&k_{12}\cdot\epsilon_{5}[\frac{1}{16}\frac{1}{s_{12}}(\epsilon_{1}\cdot\epsilon_{2})(\epsilon_{3}\cdot\epsilon_{4})]-[\frac{1}{8}\frac{1}{s_{12}}(\epsilon_{1}\cdot\epsilon_{2})(\epsilon_{3}\cdot\epsilon_{5})](k_{12}\cdot\epsilon_{4})+(k_{1}-k_{234})\epsilon_{5}[\frac{1}{16}\frac{1}{s_{23}}(\epsilon_{2}\cdot\epsilon_{3})(\epsilon_{4}\cdot\epsilon_{1})]\\
&+[\frac{1}{8}\frac{1}{s_{23}}(\epsilon_{2}\cdot\epsilon_{3})(\epsilon_{4}\cdot\epsilon_{5})](k_{234}\cdot\epsilon_{1})-\epsilon_{1}\cdot\epsilon_{5}[\frac{1}{8}\frac{1}{s_{23}}(\epsilon_{2}\cdot\epsilon_{3})(\epsilon_{4}\cdot k_{1})+\frac{1}{16s_{23}}k_{23}\cdot\epsilon_{4}\epsilon_{2}\cdot\epsilon_{3}]\\
&-\frac{s_{1234}}{16s_{12}s_{123}}k_{12}\cdot\epsilon_{3}\epsilon_{1}\cdot\epsilon_{2}\epsilon_{4}\cdot\epsilon_{5}+\frac{s_{1234}}{16s_{23}s_{234}}k_{23}\cdot\epsilon_{4}\epsilon_{2}\cdot\epsilon_{3}\epsilon_{1}\cdot\epsilon_{5}\\
&+k_{1234}\cdot\epsilon_{5}\frac{1}{16s_{12}s_{123}}k_{123}\cdot\epsilon_{4}k_{12}\cdot\epsilon_{3}\epsilon_{1}\cdot\epsilon_{2}-k_{1234}\cdot\epsilon_{5}\frac{1}{16s_{23}s_{234}}k_{234}\cdot\epsilon_{1}k_{23}\cdot\epsilon_{4}\epsilon_{2}\cdot\epsilon_{3}-\frac{1}{8s_{23}}\epsilon_{1}\cdot k_{23}\epsilon_{2}\cdot\epsilon_{3}\epsilon_{4}\cdot\epsilon_{5}\\
&-s_{1234}\sum_{12345=X\cup Y}\eta^{\alpha\beta}\eta^{\rho\sigma}\ma{T}[12345](H_{X\alpha\nu}H_{Y\mu\beta}\epsilon_{5\gamma}\tilde{\epsilon}_{5\lambda}\eta^{\gamma\mu}\eta^{\lambda\nu})\\
&+k_{1234}\cdot\epsilon_{5}\eta^{a\sigma}\sum_{12345=X\cup Y}k_{Y}^{\rho}\ma{T}[12345](H_{X\rho a}H_{Y\sigma\nu}\tilde{\epsilon}_{5\lambda}\eta^{\lambda\nu}).
\fe
One will find that some extra off-shell terms appear. The last two lines of the equation have the same origin as the 3-pt off-shell terms. In the following calculation, we will see they are the only terms left.

The extra terms from the 4-vertex terms can also be obtained. We write down the answer directly:
\ie
&\frac{1}{8}[\frac{1}{s_{12}}\epsilon_{1}\cdot\epsilon_{2}\epsilon_{3}\cdot\epsilon_{5}k_{12}\cdot\epsilon_{4}-\frac{1}{s_{23}}\epsilon_{1}\cdot k_{4}\epsilon_{2}\cdot\epsilon_{3}\epsilon_{4}\cdot\epsilon_{5}]+\frac{1}{16}[\frac{1}{s_{23}}\epsilon_{1}\cdot\epsilon_{4}\epsilon_{2}\cdot\epsilon_{3}k_{234}\cdot\epsilon_{5}]\\
&-\frac{1}{16}(\frac{1}{s_{12}}\epsilon_{1}\cdot\epsilon_{2}\epsilon_{3}\cdot\epsilon_{4}k_{12}\cdot\epsilon_{5}+\frac{1}{s_{23}}\epsilon_{2}\cdot\epsilon_{3}\epsilon_{1}\cdot\epsilon_{4}k_{1}\cdot\epsilon_{5}-\frac{1}{s_{23}}\epsilon_{2}\cdot\epsilon_{3}k_{123}\cdot\epsilon_{4}\epsilon_{1}\cdot\epsilon_{5}-\frac{1}{s_{23}}\epsilon_{2}\cdot\epsilon_{3}\epsilon_{1}\cdot\epsilon_{5}k_{1}\cdot\epsilon_{5}).
\fe
We can sum over these two parts, and finally obtain a complicated result:
\ie
&\text{OFF}_{4}:=-\frac{s_{1234}}{16s_{12}s_{123}}k_{12}\cdot\epsilon_{3}\epsilon_{1}\cdot\epsilon_{2}\epsilon_{4}\cdot\epsilon_{5}+\frac{s_{1234}}{16s_{23}s_{234}}k_{23}\cdot\epsilon_{4}\epsilon_{2}\cdot\epsilon_{3}\epsilon_{1}\cdot\epsilon_{5}\\
&+k_{1234}\cdot\epsilon_{5}\frac{1}{16s_{12}s_{123}}k_{123}\cdot\epsilon_{4}k_{12}\cdot\epsilon_{3}\epsilon_{1}\cdot\epsilon_{2}-k_{1234}\cdot\epsilon_{5}\frac{1}{16s_{23}s_{234}}k_{234}\cdot\epsilon_{1}k_{23}\cdot\epsilon_{4}\epsilon_{2}\cdot\epsilon_{3}-\frac{1}{8s_{23}}\epsilon_{1}\cdot k_{23}\epsilon_{2}\cdot\epsilon_{3}\epsilon_{4}\cdot\epsilon_{5}\\
&-s_{1234}\sum_{12345=X\cup Y}\eta^{\alpha\beta}\eta^{\rho\sigma}\ma{T}[12345](H_{X\alpha\nu}H_{Y\mu\beta}\epsilon_{5\gamma}\tilde{\epsilon}_{5\lambda}\eta^{\gamma\mu}\eta^{\lambda\nu})\\
&+k_{1234}\cdot\epsilon_{5}\eta^{a\sigma}\sum_{12345=X\cup Y}k_{Y}^{\rho}\ma{T}[12345](H_{X\rho a}H_{Y\sigma\nu}\tilde{\epsilon}_{5\lambda}\eta^{\lambda\nu})
\fe
This equation is the expression of the extra off-shell terms of the unifying relation of the 4-pt currents $\text{OFF}_{4}$. For the 4-pt B-field current, the analysis is simpler than the graviton current, and we find that the result is the same as the 4-pt graviton case:
\ie
-\ma{T}[12345]s_{1234}H_{1234ab}\epsilon_{5c}\tilde{\epsilon}_{5d}\eta^{ac}\eta^{bd}=-\ma{T}[12345]s_{1234}B_{1234ab}\epsilon_{5c}\tilde{\epsilon}_{5d}\eta^{ac}\eta^{bd}=\frac{1}{16}A_{123}\cdot\epsilon_{4}+\text{OFF}_{4}
\fe

In fact, one can obtain the extra off-shell terms for any point currents using Mathematica and finally find that the unifying relation for amplitudes can be reproduced from BG currents. However, we have not found a closed form for those extra off-shell terms yet. From the results above, we propose the following equation as an enhancement of \eqref{exp}:
\ie
\ma{T}[In]H_{Iab}\epsilon_{c}\tilde{\epsilon}_{d}\eta^{ac}\eta^{bd}=\ma{T}[In]B_{Iab}\epsilon_{c}\tilde{\epsilon}_{d}\eta^{ac}\eta^{bd}\sim A_{I}\cdot\epsilon_{n}+\text{extra off-shell terms},
\fe
which we have verified up to 4-pt currents (or rank-4) in this paper.

\section{Outlooks}
In this paper, we write down the BG currents for the extended gravity explicitly, which is equivalent to the double field theory currents. Gravity BG currents are very helpful for computing any point gravity amplitudes since this is a programming-friendly method. We also discuss the unifying relations for some low-point examples. We want to see in what sense the unifying relation is realized for BG currents and finally find that there will be many off-shell terms appearing. One can always do this analysis for any point currents if patient enough. However, the general proof is still unknown since the structure of extended gravity BG currents is extremely complicated. We hope we can solve it in the future as a corollary of the current KLT relation.

There are many problems left and we can write some of them here. 
\begin{enumerate}
\item There are lots of results for amplitudes of the Einstein-Yang-Mills (EYM) theory \cite{Nandan:2016pya,Du:2017gnh,Fu:2017uzt}. The unifying relation between the extended gravity and the EYM theory is one of them. Studying this kind of unifying relation is also helpful for finding the current KLT relation. One can also consider the expansion relation of the gravity currents as a generalization of the amplitude relation. An example is \cite{Wu:2021exa}. To find how the expansion relation is realized for gravity currents is important since it will tell us the expansion relation at the Feynman rule level rather than the amplitude level.

\item It is also interesting to reproduce the extended gravity theory, both amplitudes and BG currents, from YM theory through the double copy prescription. Similar works have been done for the double field theory and EYM \cite{Diaz-Jaramillo:2021wtl,Faller:2018vdz}.

\item It is also interesting to explore the supersymmetry case and the AdS case \cite{Zhou:2021gnu}. Some nontrivial relations have been found and the generalization to BG currents can deepen our understanding of these relations.

\item Using the sewing procedure of BG currents \cite{Chen:2023bji,Gomez:2022dzk}, the relations for the 1-loop integrands should be reproduced from BG currents. An alternative way to obtain the loop results is \cite{Lee:2022aiu}, where they exploit the Dyson-Schwinger equation to construct the off-shell recursions at an arbitrary loop level.

\item It will also be interesting to study more relations between the classical double copy \cite{Shi:2021qsb} and the extended gravity BG currents.
\end{enumerate}

\section*{Acknowledgement}
I would like to thank Yi-Jian Du for the discussions. I also want to thank Chen Huang for sharing his Mathematica code for gluon BG currents, even though I have never used it. YT is partly supported by National Key R\&D Program of China (NO. 2020YFA0713000). 
\newpage
\appendix
\section{The explicit expression of the graviton current}\label{app1}
In this appendix, we will write down the graviton current of the extended gravity theory explicitly. Before we give the concrete expression, we need to explain some new notations first. We will use the notation $H_{I\lambda\mu\nu}$ to denote the B-field strength, which is given by
\ie
H_{I\lambda\mu\nu}=ik_{I\lambda}B_{I\mu\nu}+ik_{I\mu}B_{I\nu\lambda}+ik_{I\nu}B_{I\lambda\mu}.
\fe
We will also use the notation
\ie
-s_{I}\tilde{H}_{I\mu\nu}:=&-2i\sum_{I=X\cup Y}F^{\rho\sigma}_{X}(k_{I\rho}\Gamma_{Y\mu\nu\sigma}-k_{I\nu}\Gamma_{Y\mu\rho\sigma})-2\eta^{\alpha\beta}\eta^{\rho\sigma}\sum_{I=X\cup Y}(\Gamma_{X\nu\alpha\sigma}\Gamma_{Y\mu\rho\beta}-\Gamma_{X\rho\alpha\sigma}\Gamma_{Y\mu\nu\beta})\\
&+2\eta^{\alpha\beta}\sum_{I=X\cup Y\cup Z}F^{\rho\sigma}_{X}(\Gamma_{Y\nu\rho\beta}\Gamma_{Z\mu\alpha\sigma}-\Gamma_{Y\alpha\rho\beta}\Gamma_{Z\mu\nu\sigma}+\Gamma_{Y\nu\alpha\sigma}\Gamma_{Z\mu\rho\beta}-\Gamma_{Y\rho\alpha\sigma}\Gamma_{Z\mu\nu\beta})\\
&-2\sum_{I=X\cup Y\cup Z\cup W}F_{X}^{\rho\sigma}F_{Y}^{\alpha\beta}(\Gamma_{Z\nu\alpha\sigma}\Gamma_{W\mu\rho\beta}-\Gamma_{Z\rho\alpha\sigma}\Gamma_{W\mu\nu\beta})
\fe
to denote the graviton current of the pure graviton theory.

The graviton current of the extended gravity theory is then given by
\ie
&-s_{I}H_{I\mu\nu}=-s_{I}\tilde{H}_{I\mu\nu}-\frac{1}{2} \eta^{ac}\eta^{bd}\sum_{I=X\cup Y}H_{X\mu ab}H_{Y\nu cd}+\eta^{ac}\sum_{I=X\cup Y\cup Z}F^{bd}_{X}H_{Y\mu ab}H_{Z\nu cd}\\
&-\frac{1}{2} \sum_{I=X\cup Y\cup Z\cup W}F_{X}^{ac}F_{Y}^{bd}H_{Z\mu ab}H_{W\nu cd}-\frac{1}{2}\eta^{ac}\eta^{bd}\sum_{n=1}^{\infty}\bigg[\frac{1}{n!}(\frac{-4}{D-2})^n\sum_{I=X\cup Y\cup (\bigcup_{i=1}^{n}Z_{i})}H_{X\mu ab}H_{Y\nu cd}\prod_{i=1}^{n}\phi_{Z_{i}}\bigg]\\
&+ \eta^{ac}\sum_{n=1}^{\infty}\bigg[\frac{1}{n!}(\frac{-4}{D-2})^n\sum_{I=X\cup Y\cup Z\cup(\bigcup_{i=1}^{n}W_{i})}F^{bd}_{X}H_{Y\mu ab}H_{Z\nu cd}\prod_{i=1}^{n}\phi_{W_{i}}\bigg]-\frac{2}{D-2}\sum_{I=X\cup Y}k_{X\mu}\phi_{X}k_{Y\nu}\phi_{Y}\\
&-\frac{1}{2} \sum_{n=1}^{\infty}\bigg[\frac{1}{n!}(\frac{-4}{D-2})^n\sum_{I=X\cup Y\cup Z\cup W\cup (\bigcup_{i=1}^{n}V_{i})}F^{ac}_XF^{bd}_YH_{Z\mu ab}H_{W\nu cd}\prod_{i=1}^{n}\phi_{V_{i}}\bigg]\\
&+\frac{1}{3(D-2)}\eta_{\mu\nu}\eta^{a\lambda}\eta^{bk}\eta^{cl}\sum_{I=X\cup Y}H_{Xabc}H_{Y\lambda kl}-\frac{1}{D-2}\eta_{\mu\nu}\eta^{a\lambda}\eta^{bk}\sum_{I=X\cup Y\cup Z}F^{cl}_{X}H_{Yabc}H_{Z\lambda kl}\\
&+\frac{1}{3(D-2)}\eta_{\mu\nu}\eta^{a\lambda}\eta^{bk}\eta^{cl}\sum_{n=1}^{\infty}\bigg[\frac{1}{n!}(\frac{-4}{D-2})^n\sum_{I=X\cup Y\cup (\bigcup_{i=1}^{n}Z_{i})}H_{Xabc}H_{Y\lambda kl}\prod_{i=1}^{n}\phi_{Z_{i}}\bigg]\\
&-\frac{1}{D-2}\eta_{\mu\nu}\eta^{a\lambda}\eta^{bk}\sum_{n=1}^{\infty}\bigg[\frac{1}{n!}(\frac{-4}{D-2})^n\sum_{I=X\cup Y\cup Z\cup (\bigcup_{i=1}^{n}W_{i})}F^{cl}_{X}H_{Yabc}H_{Z\lambda kl}\prod_{i=1}^{n}\phi_{W_{i}}\bigg]\\
&+\frac{1}{D-2}\eta_{\mu\nu}\eta^{a\lambda}\sum_{I=X\cup Y\cup Z\cup W}F^{bk}_XF^{cl}_YH_{Zabc}H_{W\lambda kl}-\frac{1}{3(D-2)}\eta_{\mu\nu}\sum_{I=X\cup Y\cup Z\cup W\cup V}F^{a\lambda}_XF^{bk}_YF^{cl}_ZH_{Wabc}H_{V\lambda kl}\\
&+\frac{1}{D-2}\eta_{\mu\nu}\eta^{a\lambda}\sum_{n=1}^{\infty}\bigg[\frac{1}{n!}(\frac{-4}{D-2})^n\sum_{I=X\cup Y\cup Z\cup W\cup(\bigcup_{i=1}^{n}V_{i})}F^{bk}_XF^{cl}_{Y}H_{Zabc}H_{W\lambda kl}\prod_{i=1}^{n}\phi_{V_{i}}\bigg]\\
&-\frac{1}{3(D-2)}\eta_{\mu\nu}\sum_{n=1}^{\infty}\bigg[\frac{1}{n!}(\frac{-4}{D-2})^n\sum_{I=X\cup Y\cup Z\cup W\cup V\cup(\bigcup_{i=1}^{n}U_{i})}F^{a\lambda}_XF^{bk}_YF^{cl}_ZH_{Wabc}H_{V\lambda kl}\prod_{i=1}^{n}\phi_{U_{i}}\bigg]\\
&+\frac{1}{3(D-2)}\eta^{a\lambda}\eta^{bk}\eta^{cl}\sum_{I=X\cup Y\cup Z}H_{Xabc}H_{Y\lambda kl}H_{Z\mu\nu}-\frac{1}{D-2}\eta^{a\lambda}\eta^{bk}\sum_{I=X\cup Y\cup Z\cup W}F^{cl}_{X}H_{Yabc}H_{Z\lambda kl}H_{W\mu\nu}\\
&+\frac{1}{3(D-2)}\eta^{a\lambda}\eta^{bk}\eta^{cl}\sum_{n=1}^{\infty}\bigg[\frac{1}{n!}(\frac{-4}{D-2})^n\sum_{I=X\cup Y\cup Z\cup (\bigcup_{i=1}^{n}W_{i})}H_{Xabc}H_{Y\lambda kl}H_{Z\mu\nu}\prod_{i=1}^{n}\phi_{W_{i}}\bigg]\\
&-\frac{1}{D-2}\eta^{a\lambda}\eta^{bk}\sum_{n=1}^{\infty}\bigg[\frac{1}{n!}(\frac{-4}{D-2})^n\sum_{I=X\cup Y\cup Z\cup W\cup (\bigcup_{i=1}^{n}V_{i})}F^{cl}_{X}H_{Yabc}H_{Z\lambda kl}H_{W\mu\nu}\prod_{i=1}^{n}\phi_{V_{i}}\bigg]\\
&+\frac{1}{D-2}\eta^{a\lambda}\sum_{I=X\cup Y\cup Z\cup W\cup V}F^{bk}_XF^{cl}_YH_{Zabc}H_{W\lambda kl}H_{V\mu\nu}\\
&-\frac{1}{3(D-2)}\sum_{I=X\cup Y\cup Z\cup W\cup V\cup U}F^{a\lambda}_XF^{bk}_YF^{cl}_ZH_{Wabc}H_{V\lambda kl}H_{U\mu\nu}\\
&+\frac{1}{D-2}\eta^{a\lambda}\sum_{n=1}^{\infty}\bigg[\frac{1}{n!}(\frac{-4}{D-2})^n\sum_{I=X\cup Y\cup Z\cup W\cup V\cup(\bigcup_{i=1}^{n}U_{i})}F^{bk}_XF^{cl}_{Y}H_{Zabc}H_{W\lambda kl}H_{V\mu\nu}\prod_{i=1}^{n}\phi_{U_{i}}\bigg]\\
&-\frac{1}{3(D-2)}\sum_{n=1}^{\infty}\bigg[\frac{1}{n!}(\frac{-4}{D-2})^n\sum_{I=X\cup Y\cup Z\cup W\cup V\cup U(\bigcup_{i=1}^{n}T_{i})}F^{a\lambda}_XF^{bk}_YF^{cl}_ZH_{Wabc}H_{V\lambda kl}H_{U\mu\nu}\prod_{i=1}^{n}\phi_{T_{i}}\bigg].
\fe

\bibliographystyle{JHEP}
\bibliography{exbg}

\providecommand{\href}[2]{#2}\begingroup\raggedright\begin{thebibliography}{10}

\bibitem{DeWitt:1967ub}
B.S.~DeWitt, \emph{{Quantum Theory of Gravity. 2. The Manifestly Covariant
  Theory}}, \href{https://doi.org/10.1103/PhysRev.162.1195}{\emph{Phys. Rev.}
  {\bfseries 162} (1967) 1195}.

\bibitem{DeWitt:1967uc}
B.S.~DeWitt, \emph{{Quantum Theory of Gravity. 3. Applications of the Covariant
  Theory}}, \href{https://doi.org/10.1103/PhysRev.162.1239}{\emph{Phys. Rev.}
  {\bfseries 162} (1967) 1239}.

\bibitem{DeWitt:1967yk}
B.S.~DeWitt, \emph{{Quantum Theory of Gravity. 1. The Canonical Theory}},
  \href{https://doi.org/10.1103/PhysRev.160.1113}{\emph{Phys. Rev.} {\bfseries
  160} (1967) 1113}.

\bibitem{Maldacena:1997re}
J.M.~Maldacena, \emph{{The Large N limit of superconformal field theories and
  supergravity}}, \href{https://doi.org/10.4310/ATMP.1998.v2.n2.a1}{\emph{Adv.
  Theor. Math. Phys.} {\bfseries 2} (1998) 231}
  [\href{https://arxiv.org/abs/hep-th/9711200}{{\ttfamily hep-th/9711200}}].

\bibitem{Witten:1998qj}
E.~Witten, \emph{{Anti-de Sitter space and holography}},
  \href{https://doi.org/10.4310/ATMP.1998.v2.n2.a2}{\emph{Adv. Theor. Math.
  Phys.} {\bfseries 2} (1998) 253}
  [\href{https://arxiv.org/abs/hep-th/9802150}{{\ttfamily hep-th/9802150}}].

\bibitem{Bern:2019prr}
Z.~Bern, J.J.~Carrasco, M.~Chiodaroli, H.~Johansson and R.~Roiban, \emph{{The
  Duality Between Color and Kinematics and its Applications}},
  \href{https://arxiv.org/abs/1909.01358}{{\ttfamily 1909.01358}}.

\bibitem{Berends:1987me}
F.A.~Berends and W.T.~Giele, \emph{{Recursive Calculations for Processes with n
  Gluons}}, \href{https://doi.org/10.1016/0550-3213(88)90442-7}{\emph{Nucl.
  Phys. B} {\bfseries 306} (1988) 759}.

\bibitem{Cho:2021nim}
K.~Cho, K.~Kim and K.~Lee, \emph{{The off-shell recursion for gravity and the
  classical double copy for currents}},
  \href{https://doi.org/10.1007/JHEP01(2022)186}{\emph{JHEP} {\bfseries 01}
  (2022) 186} [\href{https://arxiv.org/abs/2109.06392}{{\ttfamily
  2109.06392}}].

\bibitem{Kawai:1985xq}
H.~Kawai, D.C.~Lewellen and S.H.H.~Tye, \emph{{A Relation Between Tree
  Amplitudes of Closed and Open Strings}},
  \href{https://doi.org/10.1016/0550-3213(86)90362-7}{\emph{Nucl. Phys. B}
  {\bfseries 269} (1986) 1}.

\bibitem{Bern:2010yg}
Z.~Bern, T.~Dennen, Y.-t.~Huang and M.~Kiermaier, \emph{{Gravity as the Square
  of Gauge Theory}},
  \href{https://doi.org/10.1103/PhysRevD.82.065003}{\emph{Phys. Rev. D}
  {\bfseries 82} (2010) 065003}
  [\href{https://arxiv.org/abs/1004.0693}{{\ttfamily 1004.0693}}].

\bibitem{Mizera:2018jbh}
S.~Mizera and B.~Skrzypek, \emph{{Perturbiner Methods for Effective Field
  Theories and the Double Copy}},
  \href{https://doi.org/10.1007/JHEP10(2018)018}{\emph{JHEP} {\bfseries 10}
  (2018) 018} [\href{https://arxiv.org/abs/1809.02096}{{\ttfamily
  1809.02096}}].

\bibitem{Gomez:2021shh}
H.~Gomez and R.L.~Jusinskas, \emph{{Multiparticle Solutions to
  Einstein\textquoteright{}s Equations}},
  \href{https://doi.org/10.1103/PhysRevLett.127.181603}{\emph{Phys. Rev. Lett.}
  {\bfseries 127} (2021) 181603}
  [\href{https://arxiv.org/abs/2106.12584}{{\ttfamily 2106.12584}}].

\bibitem{Cho:2022faq}
K.~Cho, K.~Kim and K.~Lee, \emph{{Perturbations of general relativity to all
  orders and the general n$^{th}$ order terms}},
  \href{https://doi.org/10.1007/JHEP03(2023)112}{\emph{JHEP} {\bfseries 03}
  (2023) 112} [\href{https://arxiv.org/abs/2209.11424}{{\ttfamily
  2209.11424}}].

\bibitem{Gross:1986mw}
D.J.~Gross and J.H.~Sloan, \emph{{The Quartic Effective Action for the
  Heterotic String}},
  \href{https://doi.org/10.1016/0550-3213(87)90465-2}{\emph{Nucl. Phys. B}
  {\bfseries 291} (1987) 41}.

\bibitem{Cheung:2017ems}
C.~Cheung, C.-H.~Shen and C.~Wen, \emph{{Unifying Relations for Scattering
  Amplitudes}}, \href{https://doi.org/10.1007/JHEP02(2018)095}{\emph{JHEP}
  {\bfseries 02} (2018) 095}
  [\href{https://arxiv.org/abs/1705.03025}{{\ttfamily 1705.03025}}].

\bibitem{Tao:2022nqc}
Y.-X.~Tao and Q.~Chen, \emph{{A type of unifying relation in (A)dS spacetime}},
  \href{https://doi.org/10.1007/JHEP02(2023)030}{\emph{JHEP} {\bfseries 02}
  (2023) 030} [\href{https://arxiv.org/abs/2210.15411}{{\ttfamily
  2210.15411}}].

\bibitem{Chen:2023bji}
Q.~Chen and Y.-X.~Tao, \emph{{Differential operators and unifying relations for
  1-loop Feynman integrands from Berends-Giele currents}},
  \href{https://doi.org/10.1007/JHEP08(2023)038}{\emph{JHEP} {\bfseries 08}
  (2023) 038} [\href{https://arxiv.org/abs/2301.08043}{{\ttfamily
  2301.08043}}].

\bibitem{Nandan:2016pya}
D.~Nandan, J.~Plefka, O.~Schlotterer and C.~Wen, \emph{{Einstein-Yang-Mills
  from pure Yang-Mills amplitudes}},
  \href{https://doi.org/10.1007/JHEP10(2016)070}{\emph{JHEP} {\bfseries 10}
  (2016) 070} [\href{https://arxiv.org/abs/1607.05701}{{\ttfamily
  1607.05701}}].

\bibitem{Du:2017gnh}
Y.-J.~Du, B.~Feng and F.~Teng, \emph{{Expansion of All Multitrace Tree Level
  EYM Amplitudes}}, \href{https://doi.org/10.1007/JHEP12(2017)038}{\emph{JHEP}
  {\bfseries 12} (2017) 038}
  [\href{https://arxiv.org/abs/1708.04514}{{\ttfamily 1708.04514}}].

\bibitem{Fu:2017uzt}
C.-H.~Fu, Y.-J.~Du, R.~Huang and B.~Feng, \emph{{Expansion of
  Einstein-Yang-Mills Amplitude}},
  \href{https://doi.org/10.1007/JHEP09(2017)021}{\emph{JHEP} {\bfseries 09}
  (2017) 021} [\href{https://arxiv.org/abs/1702.08158}{{\ttfamily
  1702.08158}}].

\bibitem{Wu:2021exa}
K.~Wu and Y.-J.~Du, \emph{{Off-shell extended graphic rule and the expansion of
  Berends-Giele currents in Yang-Mills theory}},
  \href{https://doi.org/10.1007/JHEP01(2022)162}{\emph{JHEP} {\bfseries 01}
  (2022) 162} [\href{https://arxiv.org/abs/2109.14462}{{\ttfamily
  2109.14462}}].

\bibitem{Diaz-Jaramillo:2021wtl}
F.~Diaz-Jaramillo, O.~Hohm and J.~Plefka, \emph{{Double field theory as the
  double copy of Yang-Mills theory}},
  \href{https://doi.org/10.1103/PhysRevD.105.045012}{\emph{Phys. Rev. D}
  {\bfseries 105} (2022) 045012}
  [\href{https://arxiv.org/abs/2109.01153}{{\ttfamily 2109.01153}}].

\bibitem{Faller:2018vdz}
J.~Faller and J.~Plefka, \emph{{Positive helicity Einstein-Yang-Mills
  amplitudes from the double copy method}},
  \href{https://doi.org/10.1103/PhysRevD.99.046008}{\emph{Phys. Rev. D}
  {\bfseries 99} (2019) 046008}
  [\href{https://arxiv.org/abs/1812.04053}{{\ttfamily 1812.04053}}].

\bibitem{Zhou:2021gnu}
X.~Zhou, \emph{{Double Copy Relation in AdS Space}},
  \href{https://doi.org/10.1103/PhysRevLett.127.141601}{\emph{Phys. Rev. Lett.}
  {\bfseries 127} (2021) 141601}
  [\href{https://arxiv.org/abs/2106.07651}{{\ttfamily 2106.07651}}].

\bibitem{Gomez:2022dzk}
H.~Gomez, R.~Lipinski~Jusinskas, C.~Lopez-Arcos and A.~Quintero~Velez,
  \emph{{One-Loop Off-Shell Amplitudes from Classical Equations of Motion}},
  \href{https://doi.org/10.1103/PhysRevLett.130.081601}{\emph{Phys. Rev. Lett.}
  {\bfseries 130} (2023) 081601}
  [\href{https://arxiv.org/abs/2208.02831}{{\ttfamily 2208.02831}}].

\bibitem{Lee:2022aiu}
K.~Lee, \emph{{Quantum off-shell recursion relation}},
  \href{https://doi.org/10.1007/JHEP05(2022)051}{\emph{JHEP} {\bfseries 05}
  (2022) 051} [\href{https://arxiv.org/abs/2202.08133}{{\ttfamily
  2202.08133}}].

\bibitem{Shi:2021qsb}
C.~Shi and J.~Plefka, \emph{{Classical double copy of worldline quantum field
  theory}}, \href{https://doi.org/10.1103/PhysRevD.105.026007}{\emph{Phys. Rev.
  D} {\bfseries 105} (2022) 026007}
  [\href{https://arxiv.org/abs/2109.10345}{{\ttfamily 2109.10345}}].

\end{thebibliography}\endgroup
\end{document}